\documentstyle[12pt]{article}
\title{\bf \Huge Exact Solutions to Sourceless Charged Massive 
       Scalar Field Equation on Kerr-Newman Background }
\author{ S. Q. Wu\thanks{E-mail: emu@iopp.ccnu.edu.cn}
       and X. Cai\thanks{E-mail: xcai@wuhan.cngb.com}  \\
   \footnotesize{Institute of Particle Physics, Hua-Zhong 
		      Normal University, Wuhan, 430079, China } }

\date{\today}
\begin{document}
\maketitle
\baselineskip 30pt
\begin{quote}  

The covariant Klein-Gordon equation in the Kerr-Newman black hole geometry 
is separated into a radial part and an angular part. It is discovered 
that in the non-extreme case, these two equations belong to a generalized 
spin-weighted spheroidal wave equation. Then general exact solutions in 
integral forms and several special solutions with physical interest are 
given. While in the extreme case, the radial equation can be transformed 
into a generalized Whittaker-Hill equation. In both cases, five-term 
recurrence relations between coefficients in power series expansion of 
general solutions are presented. Finally, the connection between the radial 
equations in both cases is discussed. 

PACS number(s): 97.60.S, 04.20.J, 03.65.G, 11.10.Q
\end{quote}

\newpage
\noindent
{\large I. INTRODUCTION}

Since the Hawking effect$^1$ on black hole was found, the evaporation of 
black hole has been investigated in several coordinates by miscellaneous 
methods such as path integral approach,$^2$ tortoise coordinate ($ r_* $) 
transformation,$^{3,4}$ and $ r_* $-coordinate analytical extension,$^4$
etc. Among these methods, the generalized tortoise transformation method 
has been used widely in the discussions not only on evaporation of static 
black hole and stationary black hole, but also on that of non-stationary 
ones.$^3$ Much more progress has been made. But this method can not give 
an exact solution of the radial ($ r_* $) equation, the radial wave 
function, which can be analyzed only in asymptotic expression.

Couch$^5$ obtained a series of exact solutions by transforming the 
separated radial equation into modified Whittaker-Hill equation under 
some special conditions. But these solutions seem to have nothing to 
do with the discussion on black hole evaporation.

Solutions to generalized spheroidal wave equation have been studied to 
some extent$^{6,7}$ by using power series expansions around regular singular 
points, so that three-term recurrence relations between coefficients can 
be manipulated in terms of the continued fraction method. Leaver$^6$ has 
shown that Teukolsky master equations in Kerr geometry are, in fact, 
spin-weighted generalized spheroidal wave equations.

It appears to be more important to obtain an exact solution to the radial 
equation for this is crucial in discussing Hawking effect of black hole. 
However, it is very difficult to do so. It is this motivation that stimulates 
our present research. The main aim of this paper is to show that the separated 
radial part of a massive covariant Klein-Gordon equation on the Kerr-Newman 
black hole ( KNBH ) background is a generalized spin-weighted spheroidal wave 
equation of imaginary number order. In this paper, we shall discuss the 
solutions to a massive complex scalar field in the KNBH geometry with three 
parameters. In the non-extreme case, its general solutions of the separated 
parts are spin-weighted generalized spheroidal wave functions$^{6-8}$ and 
some special solutions to the radial equation with physical interest are 
given. General solutions to the radial equation in the extreme case shall be 
briefly discussed. Finally, we show that the radial equation in the extreme 
case is a confluent equation of that in the non-extreme case.

Sec.II deals with variable separation of a sourceless complex scalar field 
on KNBH and solutions to the angular part. In Sec.III and IV, the radial 
equation is solved in both non-extreme and extreme cases respectively. In 
III(a) we reduce the radial equation to standard form, and in III(b) and (c) 
we obtain general solutions and special ones including case ($ \omega=\mu=0 $) 
respectively. Five-term recurrences between coefficients of solutions in 
power series forms are given in both cases. In addition, we give solutions 
in integral forms and some special solutions of physical interest in the 
non-extreme case. Conditions for general solutions exist are given in these 
two cases. Sec.V is devoted to discussing the connection between the radial 
equation in the extreme case and that in the non-extreme case. Finally, we 
point out some probable applications and generalization of exact solutions 
in Sec.VI.

In appendix part, three-term recurrence relation between coefficients in 
power series expansions around regular singular points for generalized 
spheroidal wave equation are presented.

\noindent
{\large II. SEPARATION OF KLEIN-GORDON EQUATION  AND SOLUTION
	    TO THE ANGULAR EQUATION}

The Kerr-Newman line element and electromagnetic one-form are given in 
the Boyer-Lindquist coordinates as follows$^{4,9}$
\begin{eqnarray}\nonumber
 ds^2 &=& g_{\mu\nu}dx^{\mu}dx^{\nu} \\
      &=& -\frac{\Delta}{\Sigma}(dt-a\sin^2\theta d\varphi)^2
 +\frac{\sin^2\theta}{\Sigma}[adt-(r^2+a^2)d\varphi]^2
 +\Sigma(\frac{dr^2}{\Delta}+d\theta^2), \\
 {\cal{A}} &=& A_{\mu}dx^{\mu}=\frac{-er}{\Sigma}
 (dt-a\sin^2\theta d\varphi)
\end{eqnarray}

\noindent
with event horizon function $ \Delta=r^2-2Mr+a^2+e^2 $, and $ \Sigma=
r^2+a^2\cos^2\theta $, where mass $ M $, charge $ e $, specific angular 
momentum $ a=J/M $ being three parameters to describe KNBH. ( Use 
Planck units system $ G=\hbar=c=1 $, and denote $ \partial_{\mu}=
\frac{\partial}{\partial x^{\mu}} $ ).

The determinant of KNBH metric tensor is $ g={\rm det}(g_{\mu\nu})=
-\Sigma^2\sin^2\theta $, while the electromagnetic four-vector potential 
$ A_{\mu} $ apparently satisfies the following covariant Lorentz gauge 
condition:
\begin{equation}
 \nabla_{\mu}A^{\mu}=\frac{1}{\sqrt{-g}}\partial_{\mu}(\sqrt{-g}
 g^{\mu\nu}A_{\nu})=0.
\end{equation}

In curved spacetime, a sourceless scalar field $ \Phi $ with mass $ \mu $ 
and charge $ q $ obeys the covariant Klein-Gordon equation ( KGE ):
\begin{equation}
 (\Box_c-{\mu}^2)\Phi=0,
\end{equation}

\noindent
where d' Alembert operator $ \Box_c $ on KNBH background is given by
\begin{eqnarray}\nonumber
 \Box_c &\equiv& \frac{1}{\sqrt{-g}}D_{\mu}(\sqrt{-g}g^{\mu\nu}D_{\nu}) \\
\nonumber &=& \frac{1}{\Sigma}\{\frac{-1}{\Delta}[(r^2+a^2)\partial_t
 +a\partial_{\varphi}+iqer]^2+\partial_r(\Delta\partial_r) \\
 && +(a\sin\theta\partial_t+\frac{1}{\sin\theta}\partial_{\varphi})^2+
 \frac{1}{\sin\theta}\partial_{\theta}(\sin\theta\partial_{\theta})\},
\end{eqnarray}

\noindent
here covariant gauge differential operator being $ D_{\mu}=\partial_{\mu}
-iqA_{\mu}. $

The scalar wave function $ \Phi $ for KGE of Eq.(4) has a solution of
variables separable form $ \Phi(t,r,\theta,\varphi)=R(r)S(\theta)e^{i(m
\varphi-{\omega}t)} $:$^9$
\begin{eqnarray}\nonumber
 \frac{1}{\Delta}[\omega(r^2+a^2)-qer-ma]^2\Phi+\partial_r(\Delta
 \partial_r\Phi)-{\mu}^2\Sigma\Phi  \\
 -(a\omega\sin\theta-\frac{m}{\sin\theta})^2\Phi+\frac{1}{\sin\theta}
 \partial_{\theta}(\sin\theta\partial_{\theta}\Phi)=0.
\end{eqnarray}

The separated results of the above equation are
\begin{equation}
 \partial_r[\Delta\partial_rR(r)]+\{\frac{[\omega(r^2+a^2)-qer-ma]^2}
 {\Delta}-{\mu}^2(r^2+a^2)-\lambda+2ma\omega\}R(r)=0
\end{equation}

\noindent
and
\begin{equation}
 \frac{1}{\sin\theta}\partial_{\theta}[\sin\theta\partial_{\theta}S(\theta)]
 +[\lambda-\frac{m^2}{\sin^2\theta}+({\mu}^2-{\omega}^2)a^2\sin^2\theta]
 S(\theta)=0
\end{equation}

\noindent
where $ \lambda $ is a separation constant.

The general solutions to the angular part are ordinary spheroidal angular 
wave functions$^{8,10}$ with spin-weight $ s=0 $. When $ a^2({\omega}^2
-{\mu}^2)=0 $, these solutions degenerate to Legendre spherical functions.

Let $ x=\cos\theta, S(\theta)=S(x)=(1-x^2)^{m/2}\Theta(x) $, Eq.(8) should 
take the following forms:
\begin{equation}
 (1-x^2)S^{\prime\prime}(x)-2xS^{\prime}(x)+[\lambda-\frac{m^2}{1-x^2}
 +a^2({\omega}^2-{\mu}^2)(x^2-1)]S(x)=0
\end{equation}

\noindent
and
\begin{equation}
 (1-x^2)\Theta^{\prime\prime}(x)-2(1+m)x\Theta^{\prime}(x)+
 [\lambda-m(m+1)+a^2({\omega}^2-{\mu}^2)(x^2-1)]\Theta(x)=0.
\end{equation}

\noindent                            
Here and after, $ S^{\prime}(x)=\partial S(x)/ \partial x $, etc..

The eigenfunctions to Eqs.(9) and (10) are generalized spheroidal wave 
functions$^{6-8} \\ S(x)=S_{\ell}^{m,0}(c,x) $ with eigenvalue $ \lambda=
\lambda_{m\ell}+c^2 $, $ c^2=a^2({\mu}^2-{\omega}^2) $. When $ \mu=0 $, 
Eqs.(9) and (10) are special cases ($ s=0 $) of the following spin-weighted 
spheroidal wave equations$^{6-8,10} $
\begin{equation}
 (1-x^2)P^{\prime\prime}(x)-2xP^{\prime}(x)+[a^2\omega^2x^2-2a\omega
 sx-\frac{(m+sx)^2}{1-x^2}+s+\lambda^{\prime}]P(x)=0
\end{equation}

\noindent
and
\begin{equation}
 (1-x^2)Q^{\prime\prime}(x)-2[s+(1+m)x]Q^{\prime}(x)+[\lambda^{\prime}
 -(m-s)(m+s+1)-2a{\omega}sx+a^2\omega^2x^2]Q(x)=0
\end{equation}

\noindent
where $ P(x)=(1-x)^{|m+s|/2}(1+x)^{|m-s|/2}Q(x) $ and $ x=\cos\theta $.

When $ a\omega=0 $, the solutions to the above equations are Jacobi 
ultra-sphere D-functions$^{10} D_{m,s}^{\ell}(x) $ or spin-weighted 
spherical harmonic functions$^{11}$ with eigenvalue $ \lambda^{\prime}
=\ell(\ell+1)-s(s+1) $, $ \ell={\rm max}(|m|,|s|) $. In general case, the 
solutions should be the generalized spin-weighted spheroidal wave 
functions$^{6-8} P(x)=P_{m,s}^{\ell}(c,x) $, $ c^2=-a^2\omega^2 $. By 
taking account of some reasonable boundary conditions, these solutions 
could be a set of orthogonal polynomials.

In the following, we shall assume that all parameters, $ M, e, a, \mu, q, 
m $, are nonzero, and discuss the radial equation of Eq.(7) according to 
two cases, namely the non-extreme case ($ M^2\not=a^2+e^2 $) and the 
extreme case ($ M^2=a^2+e^2 $). Special case ($ \omega=\mu=0 $) will be 
included in subsection III(c).

\noindent
{\large III. SOLUTIONS TO THE RADIAL EQUATION IN THE
	     NON-EXTREME CASE ($ M^2 \not=a^2+e^2 $) }

{\large (a) Simplification of the radial equation in the case 
	($ \varepsilon\not=0 $) }

In this case, we put $ \varepsilon=\sqrt{M^2-a^2-e^2} $, ($ 0< \varepsilon 
< M $). After making substitutions of $ r=M+\varepsilon z $ and $ R(r)=
R(z)=(z-1)^{i|B+A|/2}(z+1)^{i|B-A|/2}F(z) $, the exterior horizon and 
interior horizon are located at points $ r_{\pm}=M\pm \varepsilon, ( 
z={\pm}1 ) $, respectively, the radial equation of Eq.(7) can be reduced 
to the following standard forms:
\begin{eqnarray}\nonumber
 (z^2-1)R^{\prime\prime}+2zR^{\prime}+[{\varepsilon}^2({\omega}^2-{\mu}^2)
 (z^2-1)+2\varepsilon(A\omega-M{\mu}^2)z+\frac{(Az+B)^2}{z^2-1} \\
 +(2{\omega}^2-{\mu}^2)(2M^2-e^2)
 -2qeM\omega-\lambda]R=0 \hskip 2.0cm ( 1<z<\infty )
\end{eqnarray}

\noindent
and
\begin{eqnarray}\nonumber
 (z^2-1)F^{\prime\prime}+2[iA+(1+iB)z]F^{\prime}+[{\varepsilon}^2
 ({\omega}^2-{\mu}^2)(z^2-1)+2\varepsilon(A\omega-M{\mu}^2)z \\
 +(2{\omega}^2-{\mu}^2)(2M^2-e^2)-2qeM\omega
 -\lambda+A^2-B^2+iB]F=0 \hskip 0.1cm ( 1<z<\infty )
\end{eqnarray}

\noindent
where $ A=2M\omega-qe, \varepsilon B=\omega(2M^2-e^2)-qeM-ma $.

In order to study behaviors of solutions to Eqs.(13) and (14) in the 
interval ($ -1<z<1 $), we rotate firstly $ T $ from real axis to imaginary 
axis $ T=i\tau $ after making substitution of $ z=\cosh T=\cosh({i\tau})
=\cos\tau $, then return to real z-axis $ z=\cos\tau $. Therefore, Eqs.(13) 
and (14) have corresponding forms in the interval ($ |z|<1 $) as follows
\begin{eqnarray}\nonumber
 (1-z^2)R^{\prime\prime}-2zR^{\prime}+[{\varepsilon}^2({\omega}^2-{\mu}^2)
 (z^2-1)+2\varepsilon(A\omega-M{\mu}^2)z-\frac{(Az+B)^2}{1-z^2}\\
 +(2{\omega}^2-{\mu}^2)(2M^2-e^2)-2qeM\omega-\lambda]R=0
 \hskip 2.0cm ( -1<z<1 )
\end{eqnarray}

\noindent
and
\begin{eqnarray}\nonumber
 (1-z^2)G^{\prime\prime}-2[A+(1+B)z]G^{\prime}+[{\varepsilon}^2
 ({\omega}^2-{\mu}^2)(z^2-1)+2\varepsilon(A\omega-M{\mu}^2)z \\
 +(2{\omega}^2-{\mu}^2)(2M^2-e^2)-2qeM\omega-\lambda+A^2-B^2-B]G=0
 \hskip 0.2cm ( |z|<1 )
\end{eqnarray}

\noindent
where we have made a function transformation 
$$ R(z)=(1-z)^{\frac{|B+A|}{2}}(1+z)^{\frac{|B-A|}{2}}G(z). $$

The essence of our manipulation is that we extend the domain of $ z $ 
from the real axis to complex $ z $-plane $ z=x+iy $ at first and then  
make an analytical extension on complex $ z $-plane from region outside 
unit circle ($ |z|>1 $) to that inside it ($ |z|<1 $). The handled issue 
is that only derivative terms change a negative sign, while the 
non-derivative terms, namely terms in square brackets, make no change 
in symbol. This method is equivalent to that Eqs.(13-16) are solved 
initially and then the solutions are made analytical extension on the 
complex $ z $-plane.

Both Eqs.(13) and (14) are generalized spin-weighted spheroidal wave 
equation$^{6-8}$ with imaginary number order, while both Eqs.(15) and 
(16) with real number order. The formers are suitable especially to study 
problems about scattering state, whereas the latters more convenient to 
investigate energy levels of bound states. Furthermore, we can apparently 
find out connection between poles of scattering amplitudes and energy 
levels of bound states. Actually, domain in which $ z $ takes values in 
Eqs.(13) and (14) is on the $ x $-axis, while those in Eqs.(15) and 
(16) on $ y $-axis. So these equations can be thought of equivalence. 
However, to be convenient, we have made a restriction on intervals that 
$ z $ takes values of $ 1<z<\infty $ in Eqs.(13) and (14), while of 
$ |z|<1 $ in Eqs.(15) and (16).

When $ \mu=0 $ or $ M=0 $, if taking $ R_1(z_1) $ as the first solution 
to Eq.(15) in the interval of $ |z|<1 $, then $ R_2(z_2) $, the second 
one to the same equation in that of $ |z|>1 $, might be:
\begin{equation}
 R_2(z_2)=(z_2-1)^{\frac{|B+A|}{2}}(z_2+1)^{\frac{|B-A|}{2}}
 \int\limits_{-1}^{+1}e^{i\varepsilon\omega z_2z_1}(1-z_1)^
 {\frac{|B+A|}{2}}(1+z_1)^{\frac{|B-A|}{2}}R(z_1)dz_1.
\end{equation}

\noindent
Here, we have assumed that $ \omega>0 $. Integral equation of Eq.(17) 
connects irregular solution $ R_2(z_2) $ with regular solution $ R_1(z_1) $.

Comparing Eqs.(9-12) with Eqs.(13-16), especially Eqs.(11, 12) with Eqs.(15,
16), we can draw a conclusion that the separated angular and radial equations
are ordinary differential equations of the same type, generalized
spin-weighted spheroidal wave equations.$^{6-8}$ Furthermore, we discover 
that $ \varepsilon, A, B $ correspond to $ -a, -s, -m $ respectively when 
$ \mu=0 $. There may exist three pairs of power series solutions to 
generalized spheroidal wave equations around singular points $ z=\pm 1, 
\infty $ respectively. Added with some proper boundary conditions, these 
power series expansions of spheroidal wave functions can be cut off to be 
polynomials.

Therefore, in the following subsection, we shall only study the generalized
spheroidal wave equation. The reader who has more interest in this equation, 
can find more information in Refs. 6 and 7 ( and references cited therein ).

{\large (b) General solutions to the radial equation 
	($ \varepsilon \not=0 $) }

The standard generalized spin-weighted spheroidal wave equation that 
we reduce to study is as follows
\begin{equation}
 (1-z^2)W^{\prime\prime}(z)-2[\alpha+(\beta+1)z]W^{\prime}(z)
 +[{\gamma}^2(z^2-1)+2{\delta}z+\bar{\lambda}-\beta]W(z)=0,
\end{equation}

\noindent
where $ \bar{\lambda} $ is a redefined eigenvalue which could make $ W(z) $ 
finite at $ z=\pm 1 $, and region of $ z $' taking values could be the 
whole complex $ z $-plane.

(i)  For the radial equation of Eq.(16), we have
$$ \alpha=A, \beta=B, {\gamma}^2={\varepsilon}^2({\omega}^2-{\mu}^2),
 \delta=\varepsilon(A\omega-M{\mu}^2); $$

(ii)  For the angular equation of Eq.(12), we have
$$ \alpha=s, \beta=m, {\gamma}^2=a^2{\omega}^2, \delta=-a s \omega; $$

(iii)  For the angular equation of Eq.(10), we have
$$  \alpha=0, \beta=m, {\gamma}^2=a^2({\omega}^2-{\mu}^2), \delta=0. $$

The form of Eq.(18) is invariant both under Laplace-transformation and by
changing parameters $ \alpha, \beta, \gamma^2, \delta, z $ into $ -\alpha, 
\beta, \gamma^2, -\delta, -z $ respectively. Namely, $ W(z)=W(\alpha,\beta,
\gamma,\delta;z) $ satisfies the following integral equation
\begin{eqnarray} \nonumber
 \int\limits_{0}^{+\infty}e^{-tz}W(\alpha,\beta,\gamma,\delta;z)dz
 &=& W(\frac{\delta}{\gamma},-\beta,\gamma,-\alpha \gamma;\frac{t}{\gamma})\\
 &=& W(\frac{-\delta}{\gamma},-\beta,\gamma,\alpha \gamma;\frac{-t}{\gamma}),
\end{eqnarray}
\begin{eqnarray} \nonumber
 W(\alpha,\beta,\gamma,\delta;z) &=& W(-\alpha,\beta,\gamma,-\delta;-z) \\
 &=& \int\limits_{0}^{+\infty}e^{-\gamma zt}W(\frac{-\delta}{\gamma},-\beta,
 \gamma,\alpha \gamma;t)dt .
\end{eqnarray}

The above formulae are integral solutions to Eq.(18). If one knows a 
solution, then he can obtain another by integral transformations of 
Eqs.(19) and (20). It is obvious that solutions are symmetry under 
the following condition:
$$ \alpha=\delta=0, \hskip 0.5cm ( \gamma\not=0 ). $$

\noindent
This just is the case (iii). At this moment, the symmetric solutions 
are ordinary spheroidal angular wave functions.$^{10}$

Now, we consider a solution to generalized spheroidal equation of Eq.(18) 
which is in power series form in the interval of $ -1<z<1 $. According to 
the knowledge of ordinary differential equation, one can know that Eq.(18) 
has two regular singularities ($ z=\pm 1 $) and one confluently irregular 
singular point ($ z=\infty $). As $ z=0 $ is its ordinary point, we can 
make a Taylor expansion of $ W(z) $ in the vicinity of ordinary point 
($ z=0 $):
\begin{equation}
 W(z)=W_n(z)=\sum\limits_{n=0}^{\infty}a_nz^n, \hskip 0.5cm ( |z|<1 ).
\end{equation}

Substituting power series of Eq.(21) into Eq.(18), we obtain five-term
recurrence relations between coefficients as follows
$$\hskip -4.4cm (\bar{\lambda}-{\gamma}^2-\beta)a_0-2{\alpha}a_1+2a_2=0,  $$
$$\hskip -1.65cm 2{\delta}a_0+[\bar{\lambda}-{\gamma}^2-\beta-(2+2\beta)]a_1
 -4{\alpha}a_2+a_3=0,  $$
$$  {\gamma}^2a_0+2{\delta}a_1+[\bar{\lambda}-{\gamma}^2-\beta
 -2(3+2\beta)]a_2-6{\alpha} a_3+12a_4=0, $$
$$ \cdots  \cdots $$
$$\hskip -1.4cm {\gamma}^2a_{n-2}+2{\delta}a_{n-1}
 +[\bar{\lambda}-{\gamma}^2-\beta-n(n+1+2\beta)]a_n $$ 
$$ -2(n+1){\alpha}a_{n+1}+(n+2)(n+1)a_{n+2}=0. $$

Redefine coefficients:
$$ A_n=\frac{{\gamma}^2}{\bar{\lambda}-{\gamma}^2-\beta-n(n+1+2\beta)},$$
$$ B_n=\frac{2\delta}{\bar{\lambda}-{\gamma}^2-\beta-n(n+1+2\beta)},   $$
$$ C_n=\frac{-2(n+1)\alpha}{\bar{\lambda}-{\gamma}^2-\beta-n(n+1+2\beta)},$$
$$ D_n=\frac{(n+2)(n+1)}{\bar{\lambda}-{\gamma}^2-\beta-n(n+1+2\beta)}.  $$

\noindent
When taking limits $ n \rightarrow \infty $, we have $ A_n, B_n, C_n 
\rightarrow 0 $, and $ D_n\rightarrow -1. $

Then, five-term recurrence relations become
\begin{equation}
  A_na_{n-2}+B_na_{n-1}+a_n+C_na_{n+1}+D_na_{n+2}=0 .
\end{equation}

After arranging coefficients $ A_n, B_n, C_n, D_n $ and making up them 
into a quasi-diagonal band matrix $ \Lambda $ and $ a_0, a_1, \cdots, a_n, 
\cdots $ into a column vector $ {\stackrel{\rightarrow}{a}}=(a_0, a_1, 
\cdots, a_n \\, \cdots) $, the above recurrence relations become an 
infinite tridiagonal matrix equation:
\begin{equation}
 \Lambda {\stackrel{\rightarrow}{a}}=h{\stackrel{\rightarrow}{a}}.
\end{equation}

The condition for solutions of Eq.(23) exist is that determinant of 
$ \Lambda $ is zero,
\begin{eqnarray}
 \rm {det}(\Lambda)=\left|
  \begin{array}{ccccccccc}
  1   & C_0 & 0   & 0   & 0   & 0   & 0   & 0 &\cdots \\
  B_1 & 1   & C_1 & D_1 & 0   & 0   & 0   & 0 &\cdots \\
  A_2 & B_2 & 1   & C_2 & D_2 & 0   & 0   & 0 &\cdots \\
  0   & A_3 & B_3 & 1   & C_3 & D_3 & 0   & 0 &\cdots \\
      &     &     &\cdots&\cdots&   &     &   &       \\
  0   &  O  & A_n & B_n & 1   & C_n & D_n & 0 &\cdots \\
      &     &     &\cdots&\cdots&   &     &   &
  \end{array}
      \right| =0.
\end{eqnarray}

\noindent
In fact, this condition could be satisfied, and we have $ {\rm det}(\Lambda) 
\rightarrow 0 \hskip 5pt {\rm when} \hskip 5pt n \rightarrow \infty $. 

Matrix equation of Eq.(23), together with determinant equation of Eq.(24) 
determines coefficients $ a_0, a_1, \cdots, a_n, \cdots $, and eigenvalue 
$ \bar{\lambda} $, hence, eigenvalue $ \bar{\lambda} $ will be a complicated 
function of $ \alpha, \beta, \gamma, \delta $, as well as $ n $. The second 
power series solution around the same point $ z=0 $ can be obtained by 
Frobenius's method. To be finite at $ z={\pm}1 $, power series $ W(z) $ 
could be truncated to be polynomial, and $ \alpha, \beta, \gamma, \delta $ 
could be integers or half-integers. While in general case, solutions to 
spin-weighted generalized spheroidal equation of Eq.(18) are transcendental 
functions.$^{6,7}$ 

Absolutely, solution $ W_n(z)=W_n(\alpha,\beta,\gamma,\delta;z) $ of Eq.(18) 
can be orthonormalized to constitute a set of complete functions.
\begin{equation}
 \int\limits_{-1}^{1}(1-z)^{\beta+\alpha}(1+z)^{\beta-\alpha}
 W_n(z)W_{n^{\prime}}(z)dz=\delta_{n,n^{\prime}}.
\end{equation}

\noindent
Solutions $ W_n(z) $ at infinity can have asymptotic forms $ W_n(z) 
\rightarrow e^{\mp{\gamma}z}, ( z \rightarrow {\pm}\infty, \gamma>0 ) $. 
This is consistent with that the Minkowski spacetime is an asymptotic 
spacetime of the Kerr-Newmann black hole. Thus, in-going wave and out-going 
wave at infinity can take form of plane waves.

{\large (c) Special solutions to the radial equation in the case 
   ($ \varepsilon\not=0 $)  }

In this subsection, we will base our discussion upon Eq.(15), namely
\begin{eqnarray}\nonumber
 (1-z^2)R^{\prime\prime}(z)-2zR^{\prime}(z)+[{\gamma}^2(z^2-1)
 +2{\delta}z-\frac{(\beta+{\alpha}z)^2}{z^2-1} \\
 +(2{\omega}^2-{\mu}^2)(2M^2-e^2)-2qeM
 \omega-\lambda]R(z)=0,\hskip 0.2cm ( |z|<1 )
\end{eqnarray}

\noindent
where
$$ {\gamma}^2={\varepsilon}^2({\omega}^2-{\mu}^2), \delta=\varepsilon
(A\omega-M{\mu}^2), $$
$$\alpha=A=2M\omega-qe, \beta=B=\frac{\omega(2M^2-e^2)-qeM-ma}{\varepsilon}.$$

Case-1: when $ \gamma=\delta=0 $, there exist three situations:

   i)  $ \omega=\pm \mu=qe/M\not=0, ( \alpha\not=0 ); $

   ii) $ \omega=\mu=qe/M=0, ( \alpha=0 ) $ ( This case can be thought 
   as special one in the above-head case. );

   iii) $ \omega=\mu=0, qeM\not=0, ( \alpha\not=0 ). $

Solutions in situations i) and iii) are Jacobi ultra-sphere functions $ 
R(z)=P_n^{(\beta+\alpha,\beta-\alpha)}\\(z) $,$^{10}$ whereas solutions in 
situation ii) degenerate to be Legendre functions, $ R(z)=P_n^{\beta}(z) $, 
or $ Q_n^{\beta}(z) $.

Case-2: when $ M{\mu}^2=0, \omega\not=0, \delta / (\varepsilon \omega)=
\alpha $, this case has been considered in detail by E. W. Leaver.$^6$

Case-3: when $ \alpha=\delta=0, \gamma\not=0 $, Eq.(26) is an ordinary
spheroidal wave equation,$^{10}$ and its solutions are Prolate spheroidal 
angular wave functions $ R(z)=S_n^{\beta,0}(\gamma,z) $.

Obviously, all these solutions are special cases of general solutions 
$ R_n^{(\beta+\alpha,\beta-\alpha)}(\gamma,\delta;z) \\
=(1-z)^{(\beta+\alpha)/2}(1+z)^{(\beta-\alpha)/2}
W_n(\alpha,\beta,\gamma,\delta;z) $.

Solutions in case-1 will be particular important in physics to black 
hole evaporation, as scattering cross section, stationary state energy 
levels, emission coefficients of black hole radiation, etc., could be 
analytically computed at exact theoretical level by use of Jacobi 
polynomials. Furthermore, there maybe exist special symmetry in such case.

\noindent
{\large IV. SOLUTIONS TO THE RADIAL EQUATION IN THE EXTREME CASE
	   ($ M^2=a^2+e^2 $) }

In the extreme KNBH case ($ \varepsilon=0 $), we make substitution 
$ r=M(1+x) $, then event horizon is located at a single point ($ r_h=M $), 
namely $ x=0 $, hence the radial equation of Eq.(7) can be transformed into 
the following confluent equation:
\begin{eqnarray}\nonumber
 &x^2R^{\prime\prime}(x)+2xR^{\prime}(x)+[({\omega}^2-{\mu}^2)M^2x^2
 +2(A\omega-M{\mu}^2)Mx \\
 &+(A+\frac{{\cal{B}}}{x})^2+(2{\omega}^2-{\mu}^2)(2M^2-e^2)
 -2qeM\omega-\lambda]R(x)=0
\end{eqnarray}

\noindent
where $ A=2M\omega-qe, M{\cal{B}}=B\varepsilon=\omega(2M^2-e^2)-qeM-ma $.

Defining
$$\hskip 1cm C^2=M^2({\omega}^2-{\mu}^2), D=M(A\omega-M{\mu}^2), $$
$${\lambda}_e=(2{\omega}^2-{\mu}^2)(2M^2-e^2)-2qeM\omega-\lambda
 -\frac{1}{4}+A^2. $$

\noindent
and making substitutions:
$$ x=e^{i\nu\xi},\hskip 1cm R(x)=R(\xi)=e^{-i\nu\xi /2}H(\xi)  $$
then, Eq.(27) is transformed into the generalized Whittaker-Hill equation 
( GWHE )
\begin{equation}
 -{\nu}^{-2}H^{\prime\prime}(\xi)+[C^2 e^{2i\nu\xi}+2D e^{i\nu\xi}
 +2A{\cal{B}}e^{-i\nu\xi}+{\cal{B}}^2 e^{-2i\nu\xi}+\lambda_e]H(\xi)=0.
\end{equation}

Solutions of GWHE of Eq.(28) can be regarded formally as
\begin{equation}
 H(\xi)=\sum\limits_{n=-\infty}^{+\infty}g_ne^{i n \nu\xi},\hskip 1cm
 n=0,\pm 1,\pm 2,\cdots  .
\end{equation}

Substituting Eq.(29) into Eq.(28), we obtain five-term recurrence 
relations between coefficients
$$ C^2g_{n-2}+2Dg_{n-1}+(\lambda_e+n^2)g_n+2A{\cal{B}}g_{n+1}
+{\cal{B}}^2 g_{n+2}=0, $$
\begin{equation}
 E {\stackrel{\rightarrow}{g}}=h_e{\stackrel{\rightarrow}{g}}
 =\sum\limits_{n=-\infty}^{+\infty}\sum\limits_{m=n-2}^{n+2}E_{m,n}g_n,
\end{equation}

\noindent
where we have recast recurrence relations in matrix form in Eq.(30), 
and defined matrix elements
$$ E_{n,n-2}=\frac{C^2}{\lambda_e+n^2},$$
$$ E_{n,n-1}=\frac{2D}{\lambda_e+n^2}, $$
$$ E_{n,n}=1,   $$
$$ E_{n,n+1}=\frac{2A{\cal{B}}}{\lambda_e+n^2},    $$
$$ E_{n,n+2}=\frac{{\cal{B}}^2}{\lambda_e+n^2}.  $$

Condition that solutions of simultaneous equations in Eq.(30) exist is 
that determinant det($ E $) must be zero, that is
\begin{eqnarray}
       {\rm det} ( E )&=&0,\\
     {\rm det}|E-h_eI|&=&0.
\end{eqnarray}

Secular equation of Eq.(32) is a characteristic equation that determines 
the existence of periodic solutions of Eq.(29). Solutions $ R(\xi) $ could 
be functions with period $ 4\pi / \xi $. There exist four series of 
periodic functions according to period being odd or even. Eq.(27) has 
two confluently irregular singular points $ x=0, \infty $. Behaviors of 
its solutions at event horizon $ r_h=0 ( x=0 ) $ depend upon that of 
$ R(\xi) $ at $ \xi \rightarrow \pm i\infty $ ( according to $ \nu $ 
being negative number or positive number ).

\noindent
{\large V. CONNECTION BETWEEN THE RADIAL EQUATION IN 
	   NON-EXTREME CASE AND THAT IN EXTREME CASE}

In this section, we illustrate that the radial equation of Eq.(27) in the 
extreme case is a confluent form of Eq.(13) in the non-extreme case, 
and give expression to the first thermodynamic law in the extreme KNBH 
case.

After making substitutions of $ \varepsilon=M\epsilon $, $ \epsilon z=x $, 
$ \varepsilon z=Mx $, $ {\cal{B}}=\epsilon B $, $ ( 0< \epsilon <1 ) $ in 
Eq.(13), we have
$$ r=M+\varepsilon z=M(1+x), \Delta={\varepsilon}^2(z^2-1)=M^2(x^2
 -{\epsilon}^2), $$
$$ A=2M\omega-qe, \varepsilon B=M{\cal{B}}=\omega(2M^2-e^2)-qeM-ma. $$
then, Eq.(13) is equivalent to the following one in 
the non-extreme case:
\begin{eqnarray}\nonumber
 \frac{\partial}{\partial x}[(x^2-{\epsilon}^2)\frac{\partial R(x)}{\partial 
 x}]+[M^2({\omega}^2-{\mu}^2)(x^2-{\epsilon}^2)+2M(A\omega-M{\mu}^2)x \\
 +\frac{(Ax+{\cal{B}})^2}{x^2-{\epsilon}^2}+(2{\omega}^2-{\mu}^2)(2M^2-e^2)
 -2qeM\omega-\lambda]R(x)=0.
\end{eqnarray}

Eq.(33) has two regular singular points $ x=\pm \epsilon, ( z=\pm 1 ) $ 
which are located at exterior horizon and interior horizon ( Cauchy surface) 
$ r_{\pm}=M\pm \varepsilon=M(1 \pm \epsilon) $ respectively, along with 
another irregular singular point $ x=\infty $. After taking limits $ \epsilon 
\rightarrow 0, x^2-{\epsilon}^2 \rightarrow x^2 $, Eq.(33) in the non-extreme 
case tends to Eq.(27) in the extreme case. The latter has two confluently 
irregular singular points $ x=0, \infty $. The irregular singular point 
$ x=0 $ which is located at event horizon $ r_h=M $ in the extreme case is 
just one to which two irregular singular points $ x=\epsilon $ and $ 
x=-\epsilon $ in the non-extreme case concur when $ \epsilon $ or $ 
\varepsilon \rightarrow 0 $.

In the extreme KNBH case ($ M^2=a^2+e^2 $), surface gravity $ \kappa_h=0, $ 
event horizon $ r_h=M $, reduced event horizon area $ A_h=M^2+a^2=2M^2-e^2, $ 
the first thermodynamic law of extreme Kerr-Newman black hole is expressed 
as follows
\begin{equation}
 dM=\Omega_hdJ+\Phi_hde
\end{equation}

\noindent
where $ \Phi_h=(er_h)/A_h $, and $ \Omega_h=a/A_h $ are electric potential, 
angular velocity at event horizon ($ r_h=M $) respectively.

\noindent
{\large VI. CONCLUSION}

In this paper, a sourceless charged massive scalar Klein-Gordon field 
equation has been separated into the angular and radial parts. The 
separated equations are all generalized spin-weighted spheroidal wave 
equations. In the non-extreme case, we present general solutions in 
power series expansion and that of integral forms, as well as several 
special solutions with physical interest for the radial equation. These 
solutions can be orthonormalized to a set of complete functions. In 
addition, they have asymptotic behaviors of plane waves at infinity. 
On the base of these orthogonal functions or polynomials, we can expand 
wave function of a complex scalar field to a quantized Klein-Gordon field 
on the Kerr-Newman background. In the extreme case, the radial equation 
can be reduced to modified Whittaker-Hill equation. In both cases, we 
obtain five-term recurrence relations between coefficients in power 
series expansions.

At base of this work, the quantum conservation laws about Hawking 
process and probable generalization to black hole thermodynamic laws 
can be discussed further. It is anticipated that the separated parts 
of Dirac equation in the Kerr-Newman geometry could be reduced to the 
forms of generalized spheroidal wave equation.

\vskip 0.5cm
\noindent {\bf ACKNOWLEDGMENT}

S. Q. Wu is grateful to Dr. W. Li for his helps on computer. This work is 
supported partly by the NNSF and Hubei-NSF in China.

\vskip 0.5cm
\noindent
{\bf APPENDIX} 

In this appendix, we present three-term recurrence relations between
coefficients in power series expansions around regular singular points
($ z=\pm 1 $) for spin-weighted spheroidal wave equation of Eq.(18), 
namely
$$(1-z^2)W_n^{\prime\prime}(z)-2[a+(b+1)z]W_n^{\prime}(z)
 +[c^2(z^2-1)+2dz+\lambda_n-b]W_n=0,  $$
$$ \hskip 6cm (-1<z<1)  \hskip 2.5cm  (A1) $$

Eq.(A1) has two regular singular points $ z=1 $ and $ z=-1 $, with 
indices $ \rho_-=0, -a-b $ and $ \rho_+=0, a-b $ respectively. When 
$ c,d \rightarrow 0 $, $ W_n(z) $ must tend to Jacobi polynomials.

Introducing a symbol $ \epsilon=\mp 1 $, we denote these two regular
singular points $ z=\pm 1=-\epsilon $. Then, we make power series 
expansions around regular singular points $ z=-\epsilon $ respectively, 
where we have written them in an united manner:
$$ W_n(z)=e^{-cz}\sum\limits_{n=0}^{\infty}f_n(1+\epsilon z)^n.
 \hskip 5.7cm(A2) $$

Substituting the above regular solutions of Eq.(A2) into Eq.(A1), we 
obtain three-term recurrence relations between coefficients as follows
$$ \hskip -1cm (1+b-\epsilon a)f_1+[\lambda_0+2ac-b-2\epsilon(bc+c+d)]f_0=0, $$
$$ \cdots \cdots $$
$$ (n+1)(n+1+b-\epsilon a)f_{n+1}+[\lambda_n+2ac-b-2\epsilon(bc+c+d)    $$
$$ -n(n+1+2b+4\epsilon c)]f_n+4\epsilon(nc+bc+d)f_{n-1}=0.  $$

After defining coefficients,
\begin{eqnarray}\nonumber
 A_n&=&\lambda_n+2ac-b-2\epsilon(bc+c+d)-n(n+1+2b+4\epsilon c),\\ \nonumber
 B_n&=&(n+1)(n+1+b-\epsilon a), \\ \nonumber
 C_n&=&4\epsilon[(n+b)c+d],
\end{eqnarray}

\noindent
recurrence relations for first term and $ n $-th term can be written as
$$\hskip -8.45cm B_0f_1+A_0f_0=0,    $$ 
$$ B_nf_{n+1}+A_nf_n+C_nf_{n-1}=0.\hskip 5.45cm (A3) $$

Three-term recurrence relations of Eq.(A3) can handled by continued 
fraction method,$^{6,7}$ or by matrix method ( see J. W. Liu's paper in 
Ref.7 ) as we can array $ A_n, B_n, C_n $ to make up a generalized Jacobi 
tridiagonal band matrix. Similar three-term recurrence relations can also 
be obtained by expansions in the light of Jacobi polynomials, but the 
coefficients $ A_n, B_n, C_n $ will be more complicated than those presented 
here.

The second regular solutions around the same points can be easily 
obtained by Frobenius's method, and we have not presented them here. 
Irregular solutions are connected with these regular ones by 
integrals similar to those in Eqs.(19) and (20).

In order to make $ W_n(z) $ finite at $ z=\pm 1 $, $ W_n(z) $ must be 
truncated to be polynomials, then $ W_n(z) $ is orthonormalized with 
eigenvalue $ \lambda_n $ and weight $ (1-z)^{b+a}(1+z)^{b-a} $. Hence
we have
$$ \int_{-1}^{+1}(1-z)^{b+a}(1+z)^{b-a}W_m(z)W_n(z)dz=\delta_{m,n}.
\hskip 2.7cm (A4) $$

Battle-Lemari$\acute{e}$ wavelet or Daubechies' compact support 
wavelets$^{12}$ can be used in numerical computation for matrix 
equation of Eq.(A3) and to prove convergence of polynomials $ W_n(z) $, 
but we don't pursue this goal here.

\vskip 1.5cm
\begin{enumerate}
 \item  S. W. Hawking, Nature, {\bf 248}, 30 (1974); Commun. Math. Phys. 
 {\bf 43}, 199 (1975); Phys. Rev. {\bf D 13}, 191 (1976); Lee, D., Nucl. Phys. 
 {\bf B 264}, 437 (1986); S. Hawking, R. Penrose, {\sl The Nature Of Space 
 And Time}, ( Princeton, New Jersey, 1996).
 \item  B. Hartle, S. W. Hawking, Phys. Rev. {\bf D 13}, 2188 (1976).
 \item  Z. H. Li, L. Liu, Acta Physica Sinica, {\bf 46}, 1273 (1997).
 \item  T. Damour, R. Ruffini, Phys. Rev. {\bf D 14}, 332 (1976).
 \item  W. E. Couch, J. Math. Phys. {\bf 26}, 2286 (1985); {\bf 22}, 1457 
 (1981).
 \item  E. W. Leaver, J. Math. Phys. {\bf 27}, 1238 (1986); E. D. Fackerell,
 R. G. Crossman, J. Math. Phys. {\bf 18}, 1849 (1977).
 \item  J. W. Liu, J. Math. Phys. {\bf 32}, 4026 (1992); B. D. B. Figneiredo,
 M. Novello, J. Math. Phys. {\bf 34}, 3121 (1993).
 \item  D. R. Brill, P. L. Chrzanowski, C. M. Pereira, E. D. Fackerell and
 J. R. Ipser, Phys. Rev. {\bf D 5}, 1913 (1972); B. P. Jensen, J. G. Mc 
 Laughlin and A. C. Ottewill, Phys. Rev. {\bf D 51}, 5676, (1995); M. Carmeli, 
 {\sl Classical Fields: General Relativity and Gauge Theory}, ( John Wiley 
 \& Sons, 1982).
 \item  B. Carter, Phys. Rev. {\bf 174}, 1559 (1968); L. Liu, {\sl General 
 Relativity}, ( Adv. Edu. Pub., 1987).
 \item  P. M. Morse, H. Feshbach, {\sl Methods of Theoretical Physics}, ( 
 McGraw-Hill, New York, 1953); {\sl Handbook of Mathematical Functions}, 
 edited by M. Abramowitz \& I. A. Stegun, 9th version, ( Dover, New York,  
 1972).
 \item  J. N. Goldberg, A. J. Macfarlane, E. T. Newman, F. Rohrlich and
 E. C. G. Sudarshan, J. Math. Phys. {\bf 8}, 2155 (1967).
 \item  G. Battle, Commun. Math. Phys. {\bf 110}, 601 (1987); {\bf 114}, 
 93 (1988); I. Daubechies, A. Grossmann, Y. Meyer, J. Math. Phys. {\bf 27}, 
 1271 (1986).
\end{enumerate}

\end{document}